\documentclass[%
mathleft,%
final,%
]{an}
\usepackage{graphicx}
\usepackage{natbib}
\usepackage{times}
\begin{document}

\Pagespan{1}{}
\Yearpublication{2013}%
\Yearsubmission{2012}%
\Month{1}%
\Volume{334}%
\Issue{1}%
\DOI{This.is/not.aDOI}%

\title{Determining distances using asteroseismic methods}

\author{V. Silva Aguirre\inst{1,2}\fnmsep\thanks{Corresponding author: victor@phys.au.dk}, L.~Casagrande\inst{3}, S.~Basu\inst{4}, T.~L.~Campante\inst{5,6}, W.~J.~Chaplin\inst{6,1}, D.~Huber\inst{7}, A.~Miglio\inst{6}, A.~M.~Serenelli\inst{8}, \and KASC WG\#1}
\titlerunning{Asteroseismic distance determination}
\authorrunning{Silva Aguirre et al.}
\institute{
Stellar Astrophysics Centre, Department of Physics and Astronomy, Aarhus University, Ny Munkegade 120, DK-8000 Aarhus C, Denmark
\and
Max Planck Institute for Astrophysics, Karl-Schwarzschild-Str. 1, 85748, Garching bei M\"{u}nchen, Germany
\and
Research School of Astronomy \& Astrophysics, Mount Stromlo Observatory, The Australian National University, ACT 2611, Australia
\and
Department of Astronomy, Yale University, P.O. Box 208101, New Haven, CT 06520-8101, USA
\and
Centro de Astrof\'{\i}sica and Faculdade de Ci\^encias, Universidade do Porto, Rua das Estrelas, 4150-762 Porto, Portugal
\and
School of Physics and Astronomy, University of Birmingham, Birmingham, B15 2TT, UK
\and
NASA Ames Research Center, Moffett Field, CA 94035, USA
\and
Instituto de Ciencias del Espacio (CSIC-IEEC), Facultad de  Ci\`encies, Campus UAB, 08193 Bellaterra, Spain}

\received{XXXX}
\accepted{XXXX}
\publonline{XXXX}

\keywords{stars: distances, stars: oscillations, stars: fundamental parameters, techniques: photometric.}

\abstract{%
Asteroseismology has been extremely successful in determining the properties of stars in different evolutionary stages with a remarkable level of precision. However, to fully exploit its potential, robust methods for estimating stellar parameters are required and independent verification of the results is needed. In this talk, I present a new technique developed to obtain stellar properties by coupling asteroseismic analysis with the InfraRed Flux Method. Using two global seismic observables and multi-band photometry, the technique determines masses, radii, effective temperatures, bolometric fluxes, and thus distances for field stars in a self-consistent manner. Applying our method to a sample of solar-like oscillators in the {\it Kepler} field that have accurate {\it Hipparcos} parallaxes, we find agreement in our distance determinations to better than 5\%. Comparison with measurements of spectroscopic effective temperatures and interferometric radii also validate our results, and show that our technique can be applied to stars evolved beyond the main-sequence phase.}
\maketitle
\section{Introduction}
The launch of the CoRoT \citep{ab06} and \textit{Kepler} missions \citep{gill10} produced an authentic revolution in the amount and quality of data on stellar oscillations. From the thousands of light curves obtained by these space missions asteroseismology, the study of pulsations in stars, allows us to obtain accurate stellar parameters using two global oscillation observables.

The power spectrum of solar-like oscillators is modulated in frequency by a Gaussian-like envelope, from where the frequency of maximum power $\nu_{\rm{max}}$ can be readily extracted. The near-regular pattern of high overtones presents a dominant frequency spacing called the large frequency separation, $\Delta\nu$. Applying scaling relations from solar values, these two asteroseismic observables may be used to estimate stellar properties of large numbers of solar-like oscillators \citep[e.g.,][]{ds08,sb10}.

In this talk I present a new method to derive stellar parameters, including distances, in a self-consistent manner by combining seismic determinations to the \citet{lc10} implementation of the InfraRed Flux Method (IRFM). Our results are compared with Hipparcos parallaxes, high-resolution spectroscopic temperature determinations, and interferometric measurements of angular diameters. All the relevant details and initial applications of this technique can be found in \citet{vsa11,vsa12}.
\section{Obtaining asteroseismic stellar parameters}\label{parm}
The large frequency separation $\Delta\nu$ scales as the square root of the mean density \citep[e.g.,][]{ru86}, while $\nu_\mathrm{max}$ is related to the acoustic cutoff frequency of the atmosphere \citep[e.g.,][]{tb91,kb95}. Based on the accurately known solar parameters, two scaling relations can be written for these quantities \citep[e.g.,][]{sh09}
\begin{equation}\label{eqn:mass} 
\frac{M}{M_\odot} \simeq \left(\frac{\nu_{\mathrm{max}}}{\nu_{\mathrm{max},\odot}}\right)^{3} \left(\frac{\Delta\nu}{\Delta\nu_\odot}\right)^{-4}\left(\frac{T_\mathrm{eff}}{T_{\mathrm{eff},\odot}}\right)^{3/2}, 
\end{equation}
\begin{equation}\label{eqn:rad} 
\frac{R}{R_\odot} \simeq \left(\frac{\nu_{\mathrm{max}}}{\nu_{\mathrm{max},\odot}}\right) \left(\frac{\Delta\nu}{\Delta\nu_\odot}\right)^{-2}\left(\frac{T_\mathrm{eff}}{T_{\mathrm{eff},\odot}}\right)^{1/2}, 
\end{equation}
where $T_{\mathrm{eff},\odot}= 5777\,\rm K$, $\Delta\nu_\odot = 135.1\pm0.1\,\rm \mu Hz$ and $\nu_{\mathrm{max},\odot}= 3090\pm30\,\rm \mu Hz$ are the values observed in the Sun \citep{dh11}. If we can determine the $T_\mathrm{eff}$ value, for instance via the IRFM, these relations give a determination of stellar mass and radius for each star that is independent of evolutionary models \citep[see, e.g.,][]{am09}.

The basic idea behind the IRFM is to recover for each star its bolometric $\left(\mathcal{F}_{\rm{Bol}}\right)$ and infrared monochromatic flux $\left(\mathcal{F}_{\lambda_{\rm{IR}}}\right)$, both measured at the top of Earth's atmosphere. One then compares their ratio to that obtained from the same quantities defined on a surface element of the star, i.e.~the bolometric flux $\sigma T_\mathrm{eff}^4$ and the theoretical surface infrared monochromatic flux. Once $\mathcal{F}_{\rm{Bol}}\rm{(Earth)}$ and $T_\mathrm{eff}$ are both known, the limb-darkened angular diameter, $\theta$, is trivially obtained. In the adopted implementation, the bolometric flux was recovered using multi-band photometry, and the flux outside of these bands (i.e.,~the bolometric correction) was estimated using a theoretical model flux from the \cite{ck04} grid. Thus the method relies on the input [Fe/H] and $\log\,g$, and an iterative process in $T_\mathrm{eff}$ to match the observed $\mathcal{F}_{\rm{Bol}}$.

As an initial sample to test the accuracy of our technique, we selected the stars that have accurate {\it Hipparcos} parallaxes \citep{fv07}, which accounts for only 21 targets out of the more than 500 main-sequence and subgiant stars in which \textit{Kepler} detected oscillations in its short-cadence mode \citep{wc11}. The asteroseismic observables $\Delta\nu$ and $\nu_\mathrm{max}$ were obtained from the power spectra using the pipeline described by \citet{dh09}. To recover the bolometric flux, we used multi-band  $B_TV_T$ and $JHK_S$ photometry from the {\it Tycho2} \citep{eh00} and Two Micron All Sky Survey \citep[2MASS;][]{ms06} catalogs, respectively. The infrared monochromatic flux was derived from 2MASS $JHK_S$ magnitudes only.

To apply the IRFM the metallicity of the targets must be given as an input. We have chosen the chemical composition of the targets in the following order of preference, according to availability: the latest revision of the Geneva-Copenhagen Survey \citep[GCS,][]{lc11}, spectroscopic determinations from \citet{hb12}, or the value given in the Kepler Input Catalogue \citep[KIC;][]{tb11} increased by $0.18~{\rm dex}$. The latter is the offset found between GCS and the KIC for the 11 stars common in our sample, and is similar to the $+0.21~{\rm dex}$ offset found by \citet{hb12}. Reddening must also be specified, and our calculations were made using the distance dependent three-dimensional Galactic extinction model from \citet{rd03}. The values were obtained after an iteration in distance as described by \citet{am12b}.

From the previous paragraphs it is clear that the asteroseismic method provides a mass and radius based on an input $T_\mathrm{eff}$ value, while the IRFM gives $T_\mathrm{eff}$ and the bolometric flux at a given input $\log\,g$ and [Fe/H]. In order to determine a unique set of stellar parameters for each star, we iterated the two methods in a consistent way. We started by calculating sets of IRFM effective temperatures for each star at fixed $\log\,g=2.0-5.0$ in steps of $0.5~{\rm dex}$; this translates into $T_\mathrm{eff}$ changes of less than 1\% for each $\log\,g$ step. Using $\log\,g$ determinations from the KIC as an initial guess, we interpolated in gravity and computed $T_\mathrm{eff}$ from the IRFM results. This $T_\mathrm{eff}$ value, together with $\nu_\mathrm{max}$ and $\Delta\nu$, was fed to the scaling relations to obtain a mass, radius, and thus $\log\,g$. Interpolating again in gravity gave an updated value of $T_\mathrm{eff}$, and the procedure was repeated until convergence in $\log\,g$ and $T_\mathrm{eff}$ was reached.

The 1$\sigma$ uncertainties of the parameters were obtained during the iterations. We took into account both the uncertainties in the seismic observables, as well as variations in the $T_\mathrm{eff}$ determinations arising from different photometric filters and $\log\,g$ determinations. The results are affected by the assumed value of extinction, and are very mildly dependent on the metallicity considered. To account for possible errors in reddening and composition, we have also computed sets of IRFM effective temperatures at $\log\,g=3.5$, one increasing $E(B-V)$ by $+0.01$ (the decreasing case is basically symmetric), and another one changing the metallicity by $\pm 0.1$~dex. Moreover, a Monte-Carlo simulation was run to estimate the uncertainties in $T_\mathrm{eff}$ from random photometric errors. Finally, an extra 20~K were added to the error budget to account for the uncertainty in the zero-point of the temperature scale \citep[see][]{lc10}.
\section{Results}\label{res}
\begin{figure}
\includegraphics[angle=0, width=\linewidth]{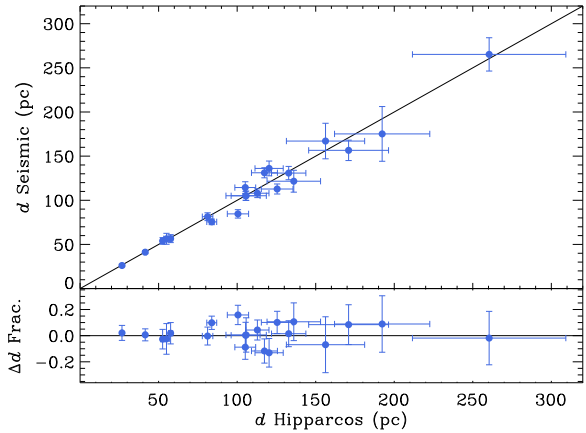}
\caption{Upper panel: comparison of {\it Hipparcos} distances with those obtained via the seismic method. Lower panels: fractional difference $({\rm Hipp.}-{\rm Seis.})$ between both determinations. Solid lines show the one-to-one correspondence.}
\label{fig:dist}
\end{figure}

Using the asteroseismic radius and the limb-darkened angular diameter $\theta$, it is straightforward to estimate the distance:
\begin{equation}
\centering
d_\mathrm{seis}=C\,\frac{2 R}{\theta}\, ,\label{eq:ang}
\end{equation}
where $C$ is the conversion factor to parsecs. In this manner, we determined asteroseismic distances for our sample targets. Fig.~\ref{fig:dist} shows the comparison between our asteroseismic distances and those obtained from {\it Hipparcos} parallax measurements. The agreement is excellent, particularly for the close-by targets, boosting our confidence on the asteroseismic parameters and the robustness of our technique. The weighted mean difference $({\rm Hipp.}-{\rm Seis.})$ is $2.3\%\pm1.8\%$.

\begin{figure}
\includegraphics[angle=0, width=\linewidth]{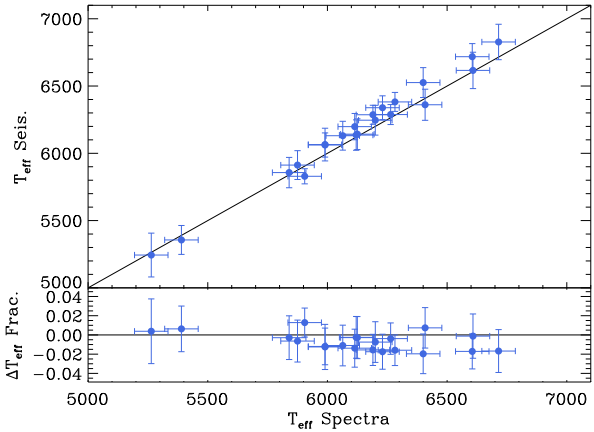}
\caption{Upper panel: comparison of effective temperatures using the asteroseismic method with those obtained from spectroscopy by \citet{hb12}. Lower panel: fractional difference $({\rm Spec.}-{\rm Seis.})$  between both determinations. Solid lines shows the one-to-one correspondence.}
\label{fig:teff}
\end{figure}

For the method to be self-consistent, we must make sure that our angular diameter and $T_\mathrm{eff}$ determinations are also robust. \citet{hb12} made a spectroscopic analysis of all but one of our targets, obtaining effective temperatures via the excitation balance of Fe~I lines at a fixed $\log g$ as determined by asteroseismology. In Fig.~\ref{fig:teff} we compare their $T_\mathrm{eff}$ values with ours and find excellent agreement. Individual fractional differences are below 2\%, while the weighted mean difference $({\rm Spec.}-{\rm Seis.})$ is $-0.8\%\pm0.4\%$. This level of agreement is particularly impressive considering that the uncertainties quoted by \citet{hb12} are of 70~K for all the targets.

\begin{figure}
\includegraphics[angle=0, width=\linewidth]{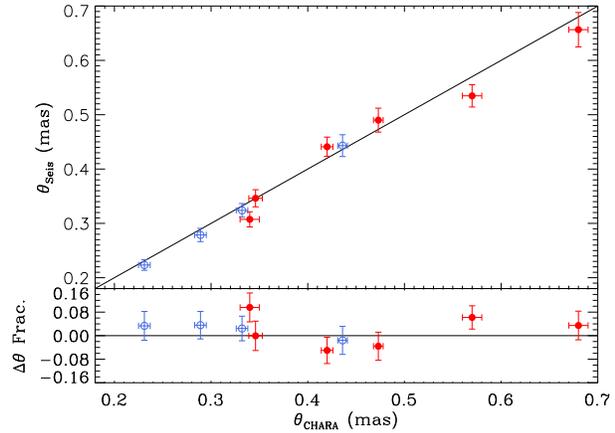}
\caption{Upper panel: angular diameters comparison of those derived by our technique to the ones obtained from spectroscopy by \citet{dh12}. Lower panel: fractional difference $({\rm CHAR.}-{\rm Seis.})$  between both determinations. Solid lines shows the one-to-one correspondence. Light open circles show the targets analyzed in this paper, while the dark filled circles depict those evolved red giants from the \citet{dh12} sample. See text for details.}
\label{fig:ang}
\end{figure}

We can verify the results of our technique by comparing our derived angular diameters with results from interferometry. In a recent observation run carried out at the PAVO/CHARA long-baseline interferometer, \citet{dh12} analyzed a total of ten asteroseismic targets, four of which are contained in our sample. We also applied our technique to the remaining six stars from the \citet{dh12} analysis, that correspond to more evolved red giants targets. In Fig.~\ref{fig:ang} we show the comparison between the angular diameters derived from interferometry and our technique, which agree remarkably well for main-sequence and red giant evolutionary phases (residual mean of $-2\%\pm2\%$). More details can be found in \citet{dh12}.

\begin{figure}
\includegraphics[angle=0, width=\linewidth]{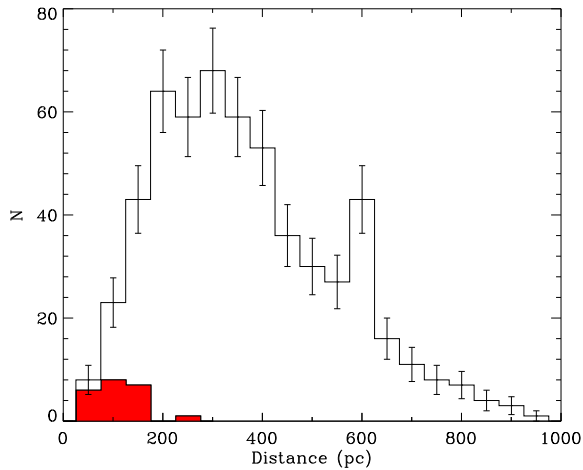}
\caption{Distance distribution derived using our technique for the complete \textit{Kepler} short-cadence sample (unshaded region). The shaded region shows the results for the 21-sample stars as a comparison.}
\label{fig:full}
\end{figure}

After verifying our technique via parallax, spectroscopic, and interferometric measurements, we applied our procedure to the full short-cadence \textit{Kepler} sample and derived consistent parameters, including distances, for all these stars. The 565 targets considered are predominantly main-sequence and subgiant stars, with a handful of red giants also present in the sample. Nevertheless, as shown in Fig.~\ref{fig:ang} our results are clearly applicable to stars evolved beyond the turnoff. All the targets have available {\it Tycho2} photometry, and we have used metallicites from the KIC increased by $0.18~{\rm dex}$. To account for the uncertainties in composition, we computed IRFM sets of $T_\mathrm{eff}$ varying the metallicity by $\pm 0.3$~dex. In Fig.~\ref{fig:full} we show a histogram with the obtained distance distribution. The results shows that we can use \textit{Kepler} data to probe populations of main-sequence and subgiant field stars as far as 1~kpc from the Sun.
\section{Conclusions}\label{conc}
Determining accurate stellar parameters is crucially important for detailed studies of individual stars, as well as for characterizing stellar populations in the Milky Way. The asteroseismic revolution produced by the {\it CoRoT} and \textit{Kepler} missions requires robust techniques to exploit fully the potential of the data and provide the community with the necessary information for ensemble analysis.

Using oscillation data and multi-band photometry, I have presented in this talk a new method to derive stellar parameters combining the IRFM with asteroseismic analysis. The novelty of our approach is that it allows us to obtain radius, mass, $T_\mathrm{eff}$, and bolometric flux for individual targets in a self-consistent manner. This naturally results in direct determinations of angular diameters and distances without resorting to parallax information, further enhancing the capabilities of our technique.

Comparison of our distance results with those from {\it Hipparcos} parallaxes shows an overall agreement better than 3\%. Furthermore, the obtained $T_\mathrm{eff}$ values show a mean difference below 1\% when compared to results from high-resolution spectroscopy.  We have also compared our calculated angular diameters with those measured by long-baseline interferometry and found agreement better than 5\%. This provides verification of our radii, $T_\mathrm{eff}$ and bolometric fluxes to an excellent level of accuracy, and shows that the results for red giant stars are also accurate.

Studies of the stellar populations in the {\it CoRoT} and {\it Kepler} fields can greatly benefit from accurate masses, radii, $T_\mathrm{eff}$, and distances \citep{am12a,am12b}. Combining this information with stellar evolutionary models and metallicity measurements can lead to an age-metallicity relation, opening the possibility of testing models of Galactic Chemical Evolution in stars outside the solar neighborhood \citep[e.g.,][]{cc97,sb09}. Applying our method to the complete short-cadence {\it Kepler} sample reveals that we can probe stars as far as 1~kpc from our Sun, making this set of main-sequence and subgiant stars extremely interesting for population studies. Although much greater distances can be probed by analyzing oscillations in giants, the ages of these stars are mostly determined by their main-sequence lifetime \citep[e.g.,][]{ms02}. Thus, the short-cadence sample is of key importance for helping to calibrate absolute mass-age relationships of red giants and correctly characterize their populations.
\acknowledgements
Funding for the {\it Kepler} Discovery mission is provided by NASA's Science Mission Directorate. The authors wish to thank the entire \emph{Kepler} team, without whom these results would not be possible. We also thank all funding councils and agencies that have supported the activities of KASC Working Group\,1. We are also grateful for support from the International Space Science Institute (ISSI). V.S.A.\ received financial support from the {\sl Excellence cluster ``Origin and Structure of the Universe''} (Garching)


\begin{thebibliography}{}
%
\bibitem[Baglin et al.(2006)]{ab06}
Baglin, A., Auvergne, M., Barge, P., et al. 2006, in ESA SP 1306, 33
%
\bibitem[Basu et al.(2010)]{sb10}  
Basu, S., Chaplin, W.~J., \& Elsworth, Y. 2010, \apj, 710, 1596
%
\bibitem[Brown et al.(1991)]{tb91} 
Brown, T. ~M., Gilliland, R.~L, Noyes, R.~W., \& Ramsey, L.~W. 1991, \apj, 368, 599
%
\bibitem[Brown et al.(2011)]{tb11} 
Brown, T. ~M.,  Latham, D.~W., Everett, M.~E, \& Esquerdo, G.~A. 2011, \aj, 142, 112
%
\bibitem[Bruntt et al.(2012)]{hb12} 
Bruntt, H., Basu, S., Smalley, B., et al. 2012, \mnras, 423, 122
%
\bibitem[Casagrande et al.(2010)]{lc10}  
Casagrande, L., Ram\'irez, I., Mel\'endez, J., et al. 2010, A\&A, 512, 54
%
\bibitem[Casagrande et al.(2011)]{lc11}  
Casagrande, L., Sch\"{o}nrich, R., Asplund, M., et al. 2011, A\&A, 530, A138
%
\bibitem[Castelli \& Kurucz(2004)]{ck04} 
Castelli, F., \& Kurucz, R.~L. 2004, arXiv:astro-ph/0405087 
%
\bibitem[Chaplin et al.(2011)]{wc11}  
Chaplin, W.~J., Kjeldsen, H., Christensen-Dalsgaard, J., et al. 2011, Science, 332, 213
%
\bibitem[Chiappini et al.(1997)]{cc97} 
Chiappini, C., Matteucci, F., \& Gratton, R. 1997, \apj, 477, 465
%
\bibitem[Drimmel et al.(2003)]{rd03}
Drimmel, R., Cabrera-Lavers, A., \& L\'opez-Corredoira, M. 2003, A\&A, 409, 205
%
\bibitem[Gilliland et al.(2010)]{gill10} 
Gilliland, R.~L., Brown, T.~M., Christensen-Dalsgaard, J., et al. 2010, \pasp, 122, 131
%
\bibitem[Hekker et al.(2009)]{sh09}  
Hekker, S., Kallinger, T., Baudin, F., et al. 2009, A\&A, 506, 465
%
\bibitem[H{\o}g et al.(2000)]{eh00}  
H{\o}g, E., Fabricius, C., Makarov, V.~V., et al. 2000, A\&A, 355, L27
%
\bibitem[Huber et al.(2009)]{dh09}  
Huber, D., Stello, D., Bedding, T.~R., et al. 2009, Comm. Aster., 160, 74
%
\bibitem[Huber et al.(2011)]{dh11}  
Huber, D., Bedding, T.~R., Stello, D. et al. 2011, \apj, 743, 143
%
\bibitem[Huber et al.(2012)]{dh12}
Huber, D., Ireland, M.~J., Bedding, T.~R., et al. 2012, \apj, in press (arXiv:1210.0012)
%
\bibitem[Kjeldsen \& Bedding(1995)]{kb95}   
Kjeldsen, H. \& Bedding, T.~R. 1995, A\&A, 293, 87
%
\bibitem[Miglio et al.(2009)]{am09}    
Miglio, A., Montalban, J., Baudin, F. et al. 2009, A\&A, 503, L21
%
\bibitem[Miglio(2012a)]{am12a}    
Miglio, A. 2012a, in {\it Red Giants as Probes of the Structure and Evolution of the Milky Way}, Astrophysics and Space Science Proceedings, ed. A. Miglio, J. Montalban, \& A. Noels (Springer-Verlag Berlin Heidelberg),p.11
%
\bibitem[Miglio et al.(2012b)]{am12b}    
Miglio, A., Chiappini, C., Morel, T. et al. 2012b, \mnras, submitted
%
\bibitem[Salaris et al.(2002)]{ms02} 
Salaris, M., Cassisi, S., \& Weiss, A. 2002, \pasp, 114, 375
%
\bibitem[Sch\"onrich \& Binney(2009)]{sb09}   
Sch\"onrich, R., \& Binney, J. 2009, \mnras, 396, 203
%
\bibitem[Silva Aguirre et al.(2011)]{vsa11}	
Silva Aguirre, V., Chaplin, W.~J., Ballot, et al. 2011, \apj, 740, L2
%
\bibitem[Silva Aguirre et al.(2012)]{vsa12}	
Silva Aguirre, V., Casagrande, L., Basu, S., et al. 2012, \apj, 757, 99
%
\bibitem[Skrutskie et al.(2006)]{ms06}  
Skrutskie, M.~,F., Cutri, R.~M., Stiening, R., et al. 2006, \aj, 131, 1163
%
\bibitem[Stello et al.(2008)]{ds08}   
Stello, D., Bruntt, H., Preston, H., \& Buzasi, D. 2008, \apj, 674, L53
%
\bibitem[Ulrich(1986)]{ru86}   
Ulrich, R.~K. 1986, \apjl, 306, L37
%
\bibitem[van Leeuwen(2007)]{fv07}
van Leeuwen, F. 2007, A\&A, 474, 653 

\end{thebibliography}
\end{document}